\documentclass{article}

\usepackage{hyperref}

\usepackage{amsmath}
\usepackage{amssymb}

\usepackage{algpseudocode}
\usepackage{algorithm}

\usepackage{natbib}

\usepackage{booktabs}

\usepackage{graphicx}

\usepackage{subcaption}

\newcommand{\argmax}{\operatornamewithlimits{argmax}}

\title{Fast Computation on Semirings Isomorphic to $(\times, \max)$ on
  $\mathbb{R}_+$}

\author{Oliver Serang \\
       Freie Universit\"at Berlin\\
       Department of Informatik\\
       Takustr. 9, 14195 Berlin, Germany /\\
       Leibniz-Institute of Freshwater Ecology and Inland Fisheries (IGB)\\
       M\"uggelsee 310, 12587 Berlin, Germany\\
       orserang@uw.edu
}

\date{\today}

\begin{document}

\sloppy

\maketitle

\begin{abstract}
Important problems across multiple disciplines involve computations on
the semiring $(\times, \max)$ (or its equivalents, the negated version
$(\times, \min)$), the log-transformed version $(+, \max)$, or the
negated log-transformed version $(+, \min)$): max-convolution,
all-pairs shortest paths in a weighted graph, and finding the largest
$k$ values in $x_i+y_j$ for two lists $x$ and $y$. However, fast
algorithms such as those enabling FFT convolution, sub-cubic matrix
multiplication, \emph{etc.}, require inverse operations, and thus
cannot be computed on semirings. This manuscript generalizes recent
advances on max-convolution: in this approach a small family of
$p$-norm rings are used to efficiently approximate results on a
nonnegative semiring. The general approach can be used to easily
compute sub-cubic estimates of the all-pairs shortest paths in a graph
with nonnegative edge weights and sub-quadratic estimates of the top
$k$ values in $x_i+y_j$ when $x$ and $y$ are nonnegative. These
methods are fast in practice and can benefit from coarse-grained
parallelization.
\end{abstract}

\section{Introduction} \label{sec:introduction}
Rings are algebraic structures in which two operations $(\otimes,
\oplus)$, generalizations of the standard $\times$ and $+$, are
supported, and where the outcomes of the operations must be found in
the set of interest (\emph{e.g.}, the ring $(\times, +)$ on the set of
integers states that adding or multiplying any two integers must yield
an integer). Importantly, the operations $\otimes$ and $\oplus$ must
be invertible: for $x$ and $y$ in the ring, $z=x \oplus y$ must be
invertible to produce either $x$ (which can be recovered as $z \oplus
-y$) or $y$ (which can be recovered as $z \oplus -x$) again (the same
is true for the operation $\times$, except the case when $x \otimes y
= 0$). The more general semirings do not necessarily include this
inverse operation. For example, on the semiring $(+, \max)$, given
only $z=\max(x, y)$ and the value of $y$, it is not possible to
retrieve the value of $x$. The greater generality of semirings makes
them important in geometry, optimization, and
physics~\cite{golan:semirings}.

The seemingly pedantic distinction between semirings and rings becomes
more pronounced when considering certain fast algorithms, which can be
applied on rings but not on semirings. A key example of this is fast
convolution with the fast Fourier transform (FFT). On the ring
$(\times, +)$, FFT can be used to perform convolution in $O(n
\log(n))$ steps (superior to the $O(n^2)$ required by a naive
convolution algorithm); however, one of the keys to the feasibility of
FFT convolution is the notion that the polynomial coefficients can be
combined into a point representation of the polynomials, which is then
operated upon, and then un-combined into the coefficients of the
product polynomial (essentially, operating on them while combined
saves substantial time). For this reason, faster-than-naive algorithms
for performing ``standard'' convolution (\emph{i.e.}, convolution on
the ring $(\times, +)$) are very well established numerical
methods~\cite{cooley:algorithm}, whereas the first algorithm with
worst-case runtime in $o(n^2)$ for max-convolution (\emph{i.e.},
convolution on the semiring $(\times, \max)$~\cite{bussieck:fast}) was
published fairly recently~\cite{bremner:necklaces} and considered by
many, including myself, to be a significant breakthrough. However,
this sub-quadratic max-convolution algorithm has a runtime that is
only slightly lower than quadratic when compared to the $O(n \log(n))$
fast standard convolution methods mentioned above. The development of
fast algorithms for max-convolution is considered important, because
FFT convolution-based dynamic programming algorithms can be used to
perform fast sum-product statistical inference on sums of two or more
random variables~\cite{tarlow2012fast, serang:probabilistic}, but when
performing max-product (\emph{i.e.}, \emph{maximum a posteriori})
inference (\emph{i.e.}, computing the best configuration) the
resulting problem is a max convolution, which was previously limited
to algorithms significantly slower than FFT~\cite{serang:fast} (with
the exception of the binary case $n=2$, wherein an $O(n \log(n))$
\emph{maximum a posteriori} algorithm based on sorting is
possible~\cite{tarlow2010hop}).

Reminiscent of the disparity between convolution over rings and
convolution over semirings is the disparity between matrix
multiplication over rings versus matrix multiplication over semirings:
Where naive matrix multiplication is in $O(n^3)$, fast matrix
multiplication once again performs operations on combined elements and
then un-combines them to achieve a runtime in $O(n^{2.807})$ with
Strassen's algorithm~\cite{strassen:gaussian}, which has since been
improved by other algorithms with the same strategy, such as with the
$O(n^{2.375})$ Coppersmith-Winograd
algorithm~\cite{coppersmith:matrix}, as well as newer variants of
improved worst-case runtime~\cite{williams:multiplying,
  le:powers}. However, the use of that un-combine step (\emph{i.e.},
inverting the $\oplus$ operation in this case) in the
faster-than-naive algorithms prevents using such algorithms on
semirings. For example, given an adjacency matrix corresponding to
some graph, matrix multiplications on the semiring $(+, min)$ can be
used to perform edge relaxations and find the shortest paths between
any two vertices in the graph. This was initially conceived as a
manner of multiplying the adjacency matrix by itself on the semiring
$(+, \min)$ to relax edges (\emph{i.e.}, when the path $u \rightarrow
x \rightarrow v$ is more efficient than the direct edge $u \rightarrow
v$, using $u \rightarrow x \rightarrow v$ as the new best distance,
which replaces the direct edge), thereby making a new adjacency matrix
containing all paths of length $\leq 2$. This process could be
repeated, with $n-1$ matrix multiplications on the semiring $(+,
\min)$, thereby yielding all most efficient paths of length $\leq n$
(equivalent to the most efficient paths overall, since no optimal path
would be longer then $n$ edges when each edge weight is in
$\mathbb{R}_+$), resulting in an algorithm that runs in
$O(n^4)$~\cite{shimbel:structural}. From there it is trivial to
achieve a speedup by computing the iterative matrix multiplication via
the powers of two:
\[
A^n =
\begin{cases} 
  {(A^2)}^{\frac{n}{2}} & n~ (\mbox{mod}~2) = 0 \\
  A \times {(A^2)}^{\frac{n-1}{2}} & else. \\
\end{cases}
\]
Matrix multiplication over the semiring $(+, \min)$ solves the
all-pairs shortest paths problem (APSP) in $\log(n)$ matrix
multiplications (or $O(n^3 \log(n))$ time); however, existing fast
matrix multiplication methods cannot be applied to the $(+, \min)$
semiring, and thus speedups substantially below $O(n^3)$ (the runtime
required to solve the APSP problem with the Floyd-Warshall
algorithm~\cite{floyd:algorithm, warshall:theorem}) are not achieved
by simply utilizing fast matrix multiplication. In fact, it has been
proven that restricting the available operations to $+$ and $<$,
implies $\Omega(n^3)$ such operations are necessary to solve the
problem exactly~\cite{kerb:effect, williams:faster}. The APSP problem
is crucially important to many different fields, including
applications where the connection to APSP is trivial (\emph{e.g.}, GPS
driving directions, routing network traffic) and applications where
the connection to APSP is nontrivial (\emph{e.g.}, a fast APSP
solution being used within the sub-quadratic max-convolution algorithm
of ~\citet{bremner:necklaces}). Research has focused primarily on
exploiting properties of particular graphs (\emph{e.g.}, using the
fact that its adjacency matrix is sparse, that the graph is planar,
\emph{etc.}), or in more general cases, approaching the problem in a
more combinatorial manner~\cite{aingworth:fast, williams:faster}.

Similar to the APSP problem, the problem of sorting all $x_i + y_j$
pairs from lists $x$ and $y$~\cite{fredman:good} exhibits a
combinatorial nature similar to the max-convolution and APSP problems;
indeed, one method for performing max-convolution is to compute the
top $k$ values in $x'_i \times y'_j$ (where $x'_i = e^{x_i}$ and $y'_j
= e^{y_j}$) and then fill them in the appropriate indices $m=i+j$ in
the max-convolution. However, despite its similarities to the two
other problems, this particular problem poses a less clear parallel
example where a fast algorithm is available for rings (rather than
semirings). All known exact approaches for this problem are in $O(n^2
\log(n))$, the same as the cost of the naive algorithm (which computes
and sorts pairs)~\cite{erickson:lower}. But the variant where only the
top $k$ values are of interest is more complicated, since the top
value is trivial (it is the maximum element in $x$ plus the maximum
element in $y$) and since the indices considered grow rapidly as $k$
increases: If the lists are first ordered so that $x_0 \geq x_1 \geq
x_2 \geq \cdots \geq x_{n-1}$ and $y_0 \geq y_1 \geq y_2 \geq \cdots
\geq y_{n-1}$, then $x_0 + y_0$ achieves the maximum, but either $x_0
+ y_1$ or $x_1 + y_0$ achieves the second highest, and the third
highest will be in $\{ x_0 + y_1, x_1 + y_0, x_0 + y_2, x_2 +
x_0\}$. Clearly, the values considered by this scheme grow in a
combinatorial manner.\newline

This paper draws upon recent work by which fast ring-based algorithms
(\emph{i.e.}, algorithms that are only appropriate for use on rings)
are used to approximate results on semirings (to which those fast
algorithms cannot be applied) via $p$-norm rings, a strategy
previously outlined for the max-convolution problem. As noted above,
until recently no known algorithms for max-convolution were even
remotely as fast as FFT convolution in practice. Here, the method for
achieving a numerical estimate of the max-convolution in $O(n
\log(n))$ time~\cite{serang:fast, pfeuffer:bounded}, is reviewed. The
exact max-convolution between two vectors $x$ and $y$ is defined as
follows:
\begin{eqnarray*}
  z &=& x ~*_{\max}~ y \\
  z_m &=& {( x ~*_{\max}~ y )}_m \\
  &=& \max_i~ x_i ~ y_{m-i} \\
  &=& \max_i~ u^{(m)}_i,\\
\end{eqnarray*}
where $u^{(m)}$ is a vector defined such that $u^{(m)}_i = x_i~
y_{m-i}$. Thus, it is possible to see the problem by first filling in
lists $u^{(0)} = (x_0 \times y_0),~ u^{(1)} = (x_0\times y_1, x_1
\times y_0), \ldots~ u^{(m)} = (x_0\times y_m, x_1 \times y_{m-1},
\ldots~ x_m \times y_0),~ \ldots$, and then the max-convolution result
(denoted $z$ above) at index $m$ will be the maximum value found in
list $u^{(m)}$. Of course, as described in this naive formulation, the
runtime is still quadratic; however it was previously noted that when
the elements of $x$ and $y$ are nonnegative, the maximum over each
vector $u^{(m)}$ could be found by exploiting the equivalence between
the maximum and the Chebyshev norm $\| \cdot \|_\infty$, and then
using that to numerically approximate the Chebyshev norm with a
$p^*$-norm, where $p^*$ is a large numerical value:
\begin{eqnarray*}
  \max_i~ u^{(m)}_i &=& \| u^{(m)} \|_\infty \\
  &=& \lim_{p \to \infty} {\left( \sum_i {\left( u^{(m)}_i \right)}^p \right)}^{\frac{1}{p}}\\
  &\approx& {\left( \sum_i {\left( u^{(m)}_i \right)}^{p^*} \right)}^\frac{1}{p^*},\\
\end{eqnarray*}
when $p^* \gg 1$~\cite{serang:fast}. When using a fixed value of $p^*$,
$u^{(m)}$ can be expanded back into its constituent pieces
\begin{eqnarray*}
  {\left( \sum_i {\left( u^{(m)}_i \right)}^{p^*} \right)}^\frac{1}{p^*} & = & {\left( \sum_i x_m^{p^*} ~y_{m-i}^{p^*} \right)}^\frac{1}{p^*} \\
  & = & {\left( x^{p^*} ~*~ y^{p^*} \right)}_m^\frac{1}{p^*}.
\end{eqnarray*}
Thus, it is apparent that the max-convolution can be performed by
taking every element of $x_i$ to the $p^*$ and taking every $y_i$ to
the $p^*$, convolving them regularly (in $n \log(n)$ time using FFT
convolution), and then taking every element in that convolution result
to the power $\frac{1}{p^*}$. Thus, a fast approximation of
max-convolution is computed in $n \log(n)$ time (and with a very fast
runtime constant, because the algorithm can make use of efficient
existing FFT libraries).

Because of numerical instability when $p^* \gg 1$ (which is limited to
underflow if the input problem is scaled so that all $x_i \in [0,1]$
and $y_i \in [0,1]$), a piecewise approach that considered $p^* \in P
\{ 1, 2, 4, 8, \ldots p^*_{\max} \}$ was used, and the highest
numerically stable value of $p^*$ is used at each index. This is
accomplished by computing a single FFT-based max-convolution for each
$p^*$ and then at each index $m$ in the result, finding the highest
$p^*$ that produces a result at that index where ${\left( x^{p^*} ~*~
  y^{p^*} \right)}_m \geq \tau$ (where $\tau$ is the error threshold
for the convolution algorithm). Because $\| u^{(m)} \|_{p^*}$ is an
upper bound of $\| u^{(m)} \|_{\infty}$, when ${\| u^{(m)}
  \|}^{p^*}_{p^*} \geq \tau$, then ${\| u^{(m)} \|}^{p^*}_{\infty}
\geq \tau$, and thus the algorithm does not suffer critical
underflow~\cite{pfeuffer:bounded}. This achieves a stable estimate
with bounded error when $p^*_{\max} \in O(\log(n))$, and thus the full
procedure (over all $p^*$ considered) can be performed in $O(n \log(n)
\log(\log(n)))$. 

When the inputs are $\in [0,1]$, the worst-case relative error has been
bounded~\cite{pfeuffer:bounded}:

\begin{eqnarray*}
  \frac{| \| u^{(m)} \|_{p^*} - \| u^{(m)} \|_\infty |}{\| u^{(m)} \|_\infty} &=& \frac{\| u^{(m)} \|_{p^*} - \| u^{(m)} \|_\infty}{\| u^{(m)} \|_\infty} \\
  &=& \frac{\| u^{(m)} \|_{p^*}}{\| u^{(m)} \|_\infty} - 1 \\
  &\leq& n^\frac{1}{p^*} - 1,
\end{eqnarray*}

where $p^*$ is the largest $p^* \in P$ that produces a numerically
stable result. By dividing the estimate by the maximum possible value
of $\frac{\| u^{(m)} \|_{p^*}}{\| u^{(m)} \|_\infty}$, (\emph{i.e.},
dividing by $n^\frac{1}{p^*}$) the worst-case relative error can be
decreased to $| n^\frac{-1}{p^*} - 1 | = 1 - n^\frac{-1}{p^*}$.

But even more significantly, rather than simply use $\| u^{(m)}
\|_{p^*}$ to approximate the maximum, ~\citet{pfeuffer:bounded} also
proposed a method for using the \emph{shape} of the $p^*$ vs. $\|
u^{(m)} \|_{p^*}$ curve to estimate $\| u^{(m)} \|_\infty$. There are
multiple ways to estimate $\| u^{(m)} \|_\infty$ from this curve, but
one way that achieves a good balance between accuracy and efficiency
is to model the norm sequence by representing the unique elements in
$u^{(m)}$ as a multiset and then projecting down to a smaller number
of elements:

\begin{eqnarray*}
\| u^{(m)} \|_{p^*}^{p^*} & = & \sum_i {u^{(m)}_i}^{p^*} \\
& = & \sum_j^{e_m} h_j {\beta_j}^{p^*},
\end{eqnarray*}

where each $\beta_j$ is one of the $e_m$ unique values in $u^{(m)}$
and $h_j$ is the number of occurences of $\beta_j$. From this
perspective, it is possible to use empirically observed values of $\|
u^{(m)} \|_{p^*}$ (for a few different $p^* \in P$) and then project
the $e_m$ unique $\beta_j$ values and their respective counts $h_j$
down to a smaller number of $r \leq e_m$ unique $\alpha_j$ values and
their respective counts $n_j$:

\begin{eqnarray*}
\| u^{(m)} \|_{p^*}^{p^*} & \approx & \sum_j^{r} n_j {\alpha_j}^{p^*}. \\
\end{eqnarray*}

Given $2 r$ norms from evenly spaced norms $\| u^{(m)} \|_{p^*}, \|
u^{(m)} \|_{2 p^*}, \| u^{(m)} \|_{3 p^*}, \ldots$, the projection
onto $r$ unique values $\alpha_1, \alpha_2, \ldots \alpha_r$ has been
shown to be zeros of the polynomial
\[
\forall j,~ \sum_{i=0}^r \gamma_i \alpha_j^{i p^*} = 0
\]
(these zeros can be found by solving for the roots of the polynomial
$\sum_{i=0}^r \gamma_i a^i$ and then taking each root to the power
$\frac{1}{p^*}$), where $\gamma_0, \gamma_1, \ldots \gamma_r$, the
coefficients of the polynomial, are defined by
\[ \left[
\begin{array}{c}
\gamma_0\\
\gamma_1\\
\gamma_2\\
\vdots\\
\gamma_r
\end{array}
\right] \in null\left(
\left[
\begin{array}{ccccc}
\| u^{(m)} \|_{p^*}^{p^*} & \| u^{(m)} \|_{2 {p^*}}^{2 {p^*}} & \| u^{(m)} \|_{3 {p^*}}^{3 {p^*}} & \cdots & \| u^{(m)} \|_{(r+1) {p^*}}^{(r+1) {p^*}} \\
\| u^{(m)} \|_{2 {p^*}}^{2 {p^*}} & \| u^{(m)} \|_{3 {p^*}}^{3 {p^*}} & \| u^{(m)} \|_{4 {p^*}}^{4 {p^*}} & \cdots & \| u^{(m)} \|_{(r+2) {p^*}}^{(r+2) {p^*}} \\
\| u^{(m)} \|_{3 {p^*}}^{3 {p^*}} & \| u^{(m)} \|_{4 {p^*}}^{4 {p^*}} & \| u^{(m)} \|_{5 {p^*}}^{5 {p^*}} & \cdots & \| u^{(m)} \|_{(r+3) {p^*}}^{(r+3) {p^*}} \\
\vdots & \vdots & \vdots & & \vdots \\
\| u^{(m)} \|_{(r) {p^*}}^{(r) {p^*}} & \| u^{(m)} \|_{(r+1) {p^*}}^{(r+1) {p^*}} & \| u^{(m)} \|_{(r+2) {p^*}}^{(r+2) {p^*}} & \cdots & \| u^{(m)} \|_{2 r {p^*}}^{2 r {p^*}} \\
\end{array}
\right]
\right).
\]
The maximum value in $u^{(m)}$ can thus be estimated as $\max \{
\alpha_1, \alpha_2, \ldots \alpha_r \}$.

When $r=1$, this projection will be of the form
\begin{eqnarray*}
\| u^{(m)} \|_\infty &\approx& {\left( \frac{ \| u^{(m)} \|_{p^*} }{ \| u^{(m)} \|_{\frac{p^*}{2}}} \right)}^\frac{1}{p^*} \\
&=& {\left( \| u^{(m)} \|_\infty \frac{ \| v \|^2_{2} }{ \| v \|_1} \right)}^\frac{1}{p^*},
\end{eqnarray*}
for some vector $v$ with elements in $[0,1]$ and where at least one
element equals $1$ (w.l.o.g., let $v_1 = 1$). Therefore, the worst-case relative
error takes the form
\[
1 - {\left( \frac{ \| v \|_2^2 }{ \| v \|_1} \right)}^\frac{1}{p^*}
\]
and will be maximized when the estimate ${\left( \frac{ \| v \|_2^2 }{
    \| v \|_1} \right)}^\frac{1}{p^*}$ attains a minimum, which
corresponds to minimizing $\frac{ \| v \|_2^2 }{ \| v \|_1}$. Aside from
boundary points on the $[0,1]$ hypercube, those extrema will occur
when
\[
\nabla \frac{\| v \|_2^2}{\| v \|_1} = 0
\]
or equivalently,
\[
\forall i>1,~
\frac{\left( \frac{\partial}{\partial v_i} \| v \|_2^2 \right) \| v \|_1 - \left( \frac{\partial}{\partial v_i} \| v \|_1 \right) \| v \|_2^2}{ {\| v \|}^2} = 0.
\]
Because $v \succeq 0$ and $v_1 = 1$, then the denominator ${\| v \|}^2
> 0$, and it is therefore possible to exploit symmetry between the
equations from two different partial derivatives $j \neq i$ (where
neither $i$ nor $j$ is $1$, because $v_1=1$ is now a constant):
\begin{eqnarray*}
\frac{\left( \frac{\partial}{\partial v_i} \| v \|_2^2 \right) \| v \|_1 - \left( \frac{\partial}{\partial v_i} \| v \|_1 \right) \| v \|_2^2}{ {\| v \|}^2} &=& \frac{\left( \frac{\partial}{\partial v_i} \| v \|_2^2 \right) \| v \|_1 - \left( \frac{\partial}{\partial v_i} \| v \|_1 \right) \| v \|_2^2}{ {\| v \|}^2} \\
\left( \frac{\partial}{\partial v_i} \| v \|_2^2 \right) \| v \|_1 - \left( \frac{\partial}{\partial v_i} \| v \|_1 \right) \| v \|_2^2 &=& \left( \frac{\partial}{\partial v_i} \| v \|_2^2 \right) \| v \|_1 - \left( \frac{\partial}{\partial v_i} \| v \|_1 \right) \| v \|_2^2\\
2 v_i \| v \|_1 - \| v \|_2^2 &=& 2 v_j \| v \|_1 - \| v \|_2^2\\
v_i &=& v_j.
\end{eqnarray*}
Therefore, a critical point for $v$ contains at most two unique values
(including $v_1=1$). The value $\frac{\|v\|_2^2}{\|v\|_1}$ takes the
form $\frac{ 1 + (n-1) \lambda^2 }{ 1 + (n-1) \lambda}$. The extrema
can now be found by optimizing with respect to $\lambda$, which yields
$\lambda = \frac{\sqrt{n} - 1}{n-1}$, for which
$\frac{\|v\|_2^2}{\|v\|_1} = 2 \frac{\sqrt{n} - 1}{n-1}$. Therefore,
the worst-case relative error with the $r=1$ projection is bounded by
${\left( 2 \frac{\sqrt{n} - 1}{n-1} \right)}^\frac{1}{p^*}$.

With $r=2$, the projection likewise has a closed form (described by
\citet{pfeuffer:bounded}), and empirical evidence suggests that the
worst-case error will be achieved with three unique values in
$u^{(m)}$ (meaning there will likewise be three unique values in $v$,
including $v_1 = 1$). Although the ability to achieve the worst-case
error for $r=2$ projection using three unique values has not been
proven, if it were true, it would imply that the worst-case relative
error for the $r=2$ projection is $< 1 - {0.7}^{\frac{4}{p^*}}$,
meaning that it no longer depends on $n$.

Essentially the projection methods use the shape of the $p^*$ vs. $\|
u^{(m)} \|_{p^*}$ curve to estimate the maximum value in
$u^{(m)}$. Furthermore, when only a constant number of $p^*$ are
considered ($| P | \in O(1)$), numerical max-convolution can be
approximated numerically in $O(n \log(n))$ with a practical runtime
only slightly slower than standard FFT convolution.\newline

For the sake of simplicity, this manuscript abstracts the projection
step into a black box function $\mbox{\bf
  estimateMaxFromNormPowerSequence}$, which accepts a spectrum of
norms of some vector $u^{(m)}$ and then uses them to estimate the
maximum value in $u^{(m)}$.

This manuscript demonstrates that the fast max-convolution method
described above generalizes as a strategy for all problems on
semirings isomorphic to the semiring $(\times, \max)$ on nonnegative
values. By exploiting the fact that a sequence of $(\times, +)$
computations in different $p^*$-norms can be used to estimate the
results on the semiring $(\times, \max)$, high-quality approximations
can be achieved using off-the-shelf fast algorithms limited to
rings. In addition to the max-convolution method previously described,
the generalized approach is demonstrated on the two other difficult
semiring problems: the APSP problem and finding the top $k$ values in
$x_i + y_j$. Using a family of $O(1)$ many $p^*$-norms for $p^* \in
P$, the ring $(\times, +)$ defined for each $p^*$-norm can be solved
using faster-than-naive algorithms (which can be applied because
operations are performed in a $(\times, +)$ ring). Then, for any index
in the result, the sequence of results at each $p^*$-norm can be used
to approximate the $(\times, max)$ semiring. Using this general
approach, it is possible to compute a high-quality approximation for
the APSP problem in sub-cubic time and compute a high-quality
approximation of the top $k$ values in $x_i + y_j$ (including
estimates of their indices $(i,j)$, which cannot be computed by any
other known approximation strategies) in $O((n+k) \log(n))$.

\section{Methods} \label{sec:methods}

\subsection{A General Approach to Adapting Fast Algorithms to Nonnegative Semirings}
The outline of the approach presented here is as follows: $F$ is a
naive algorithm of interest that inefficiently solves a problem on the
semiring $(\times, \max)$. $G$ is an identical algorithm on the ring
$(\times, +)$. $H$ is an algorithm that produces an equivalent result
to $G$, but does so in a faster manner (\emph{e.g.}, FFT convolution,
Strassen matrix multiplication, \emph{etc.}). It is then demonstrated
that, for a family of $p^*$ values, $G$ can be used to compute a
sequence of $p^*$-norms, which can be used to approximate $F$. The
same approximation can therefore be achieved using $H$ instead of $G$,
since $H$ exhibits identical behavior to $G$ (even if it internally
performs computations in a manner completely different from
$G$).\newline

The semiring of interest defines $(\times, \max)$ on the real
numbers. Let $I$ be a collection of nonnegative real-valued
inputs. Let $F$ be a finite sequence of operations from the semiring
of interest (\emph{i.e.}, $\max$ and $\times$ operations) on the
values in collection $I$, which returns a new collection of those
results $F(I)$. Let $G$ be a finite sequence of operations created by
replacing every $\max$ operation in $F$ with a $+$ operation. Where
calling $F(I)$ returned some collection, now $G(I)$ will return a
collection of identical shape (\emph{i.e.}, the return values can be
thought of as tensors of the same dimension and shape). Let $H$ be a
finite sequence of operations where, for every valid input $I$, $H(I)$
returns a result numerically indistinguishable to calling
$G(I)$. Because many fast algorithms used for $H$ will have less
numerical stability than their naive counterparts, stipulate that
$H(I)_j \approx G(I)_j$ (\emph{i.e.}, the result value at some index
$j$ of the result tensor is numerically indistinguishable) whenever
the result $H(I)_j$ has experienced neither critical underflow nor
overflow. In summary, stipulate that $H(I)_j \approx G(I)_j$ whenever
$H(I)_j$ approaches neither zero nor infinity. If the means by which
overflow occurs is limited (this is achieved by scaling the inputs in
$[0,1]$, described later in this manuscript), then it is sufficient to
stipulate that $H(I)_j$ is accurate when sufficiently greater than
zero. In other words, ${G(I)}_j \approx {H(I)}_j$ when ${H(I)}_j \geq
\tau$, where $\tau$ is a value that depends on the numeric stability
of $H$.

Now consider that any sequence of operations creating an expression in
$F$ can be distributed and rearranged into an equivalent expression:
$a \times \max \{ c, d\} = \max \{ a \times c, a \times
d\}$. Therefore, any of the results computed by $F$ must be of the
form $F(I)_j = \max \{ \alpha_1, \alpha_2, \alpha_3, \ldots \}$, where
the $\alpha_i$ values are the results of finite numbers of products of
the inputs. Note that the ring $(\times, +)$ possesses a similar
property: $a \times (b + c) = (a \times b) + (a \times c)$, implying
any results from $G$ or $H$ will be equal to an expression of the form
$H(I)_j \approx G(I)_j = \alpha_1 + \alpha_2 + \alpha_3 + \cdots$
where the $\alpha_i$ values are the result of finite numbers of
products on the inputs. Because the $\alpha_1, \alpha_2, \ldots$
values are products of combinations of the inputs, then calling
$H(I^{p^*})$ (where $I^{p^*}$ denotes taking each element of $I$ to
the power $p^*$) will produce results of the form $H(I^{p^*})_j =
\alpha_1^{p^*} + \alpha_2^{p^*} + \alpha_3^{p^*} \cdots$, because
${x_1}^{p^*} \times {x_2}^{p^*} \times \cdots = {\left( x_1 \times x_2
  \times \cdots \right)}^{p^*}$.

When taking the input values to large powers $p^*$, values larger than
$1$ quickly become large and values smaller than $1$ quickly approach
zero. For this reason, let
\[ H'(I, p^*) = H\left( {\left( \frac{I}{\max_i~ I_i} \right)}^{p^*} \right) \times \max_i~ I_i~. \]
Then the error directly introduced by moving to the $p^*$-norm ring
will be limited to underflow, since all inputs will be scaled to
values in $[0,1]$. For any given result at index $j$, values of $p^*$
that result in critical levels of underflow will be excluded by
verifying that $H'(I, p^*)_j \geq \tau$. Because overflow is no longer
considered at index $j$, and because critical underflow has been ruled
out by the definitions of $H$ and $\tau$ chosen above, then it follows
that $G(I^{p^*})_j \approx H'(I, p^*)_j$. For example, when using a
high-quality FFT library for fast convolution of vectors with elements
in $[0,1]$ and of length $n$ (where $n$ is small enough to permit
storing the vectors in RAM on current computers), a result value at
some index $j$ where FFT convolution is greater than roughly
${10}^{-12}$ indicates that FFT convolution was stable to
underflow. And since significant overflow cannot have occurred, this
result is numerically stable with respect to underflow and overflow,
because overflow was eliminated by first scaling the
problem~\cite{pfeuffer:bounded}. So if $G(I)$ performs the naive
convolution between two vectors and $H(I)$ performs the FFT
convolution between the same vectors, $H(I)_j \geq \tau$ at some index
$j$ implies that $H(I)_j \approx G(I)_j$, where $\tau = {10}^{-12}$.

First compute $H(I^{p^*}),~ \forall p^* \in P$. Then, for any given
result index $j$ it is possible to produce points on the curve $\left(
p^*, H'(I, p^*)_j \right),~ \forall p^* \in P$. This curve can be
thought of as a sequence of $p^*$-norms taken to power $p^*$, where
the point paired with $p^*$ is of the form
\[ H'(I, p^*)_j = \| \left( \alpha_1, \alpha_2, \alpha_3, \ldots \right)
\|_{p^*}^{p^*}. \]

This curve can be used to compute $E_j(I)$, a numerical estimate of
the true maximum at index $j$. If the vector used to define result $j$
is denoted $u = \left( \alpha_1, \alpha_2, \alpha_3, \ldots \right)$,
then
\begin{multline*}
{F(I)}_j = \max_i~ u_i \approx\\
E_j(I) = \mbox{\bf estimateMaxFromNormPowerSequence}\left( (~
\| u \|_{p^*}^{p^*} ~\vert ~ \forall p^* \in P : \| u \|_{p^*}^{p^*} \geq
\tau ~) \right).
\end{multline*}

Using this strategy, a faster-than-naive algorithm $H$ can be called a
constant number of times ($|P| \in O(1)$) to approximate $F$. In the
case of the fast numerical approach to max-convolution described in
section~\ref{sec:introduction}, $F$ corresponds to the naive $O(n^2)$
max-convolution algorithm, which performs precisely the desired
operations in the most naive manner on the semiring $(\times,
\max)$. $G$ corresponds to a naive $O(n^2)$ standard convolution
algorithm; this naive standard convolution algorithm could be used
$|P|$ times to perform the appropriate operations in $L_{p^*}$ space
$\forall p^* \in P$, and those $p^*$-norm results can later be
aggregated to approximate the exact result from $F$. However, no
speedup is achieved when using the algorithm $G$ in this manner; in
fact, this process will almost certainly be slower, since the code of
$G$ is nearly identical to $F$ and is called a small number of times
for the different $p^*$ values, whereas $F$ is only called once. In
this case, $H$ corresponds to an $O(n \log(n))$ FFT standard
convolution algorithm, which is chosen exclusively based on its
numerical equivalence to $G$. For this reason, even if the steps in
$H$ include operations other than $\times$ or $+$ (such as with FFT
convolution), its underlying equivalence to $G$ still affords much
more efficient estimation of the various $p^*$-norms. These norms can
in turn, permit numerical estimation of $F$, even if there is no
apparent direct connection between $F$ and $H$.

\subsection{Fast Estimation of All-Pairs Shortest Paths Distances}
First, the proposed approach is demonstrated on a classic computer
science problem, the APSP problem. In the most general case, when the
adjacency matrix is dense (\emph{i.e.}, when all pairs of nodes in the
graph are joined by an edge), sub-cubic runtimes have been achieved by
complicated algorithms. The approach estimates the APSP resultant path
lengths in $O(M(n) \log(n))$ steps, where $M(n)$ is the cost of
standard floating-point matrix multiplication ($M(n)$ is $O(n^3)$ with
naive matrix multiplication, $O(n^{2.807})$ with Strassen's
multiplication method, $O(n^{2.36})$ with the Coppersmith-Winograd
algorithm, \emph{etc.}). The proposed method can be applied to graphs
(or directed graphs) with nonnegative edge weights $W_{u,v} \geq 0,
~(u,v) \in E$ (\emph{i.e.}, where $W$ describes the adjacency matrix
of the graph). The APSP problem seeks to find the shortest path
between every pair of vertices: for any two vertices $(u,v)$, the
shortest distance computed in the APSP problem would consider all
paths from $u$ to $v$ and find the one with shortest distance,
including direct paths, paths that pass through a single other vertex
$x$, paths that pass through vertices $x$ then $y$, \emph{etc.}:
$\min~ \{ W_{u,v}, \min_x~ W_{u,x} + W_{x,v}, \min_{x,y}~ W_{u,x} +
W_{x,y} + W_{y,v}, \ldots \}$.

Here, the input consists of the collection of weights $I = W$. Define
the naive algorithm to iteratively relax edges (\emph{i.e.}, it
repeatedly finds shorter edges as it progresses) by performing
$\log(n)$ matrix multiplications on the semiring $(+, \min)$. Aside
from statically bounded looping instructions (which could be unrolled
for any particular problem), each of those matrix multiplications is
constructed entirely of operations in the semiring $(+,
\min)$. Because the runtime is dominated by the $(+, \min)$ matrix
multiplications, it is possible to first simplify by letting $F$ be
the naive matrix multiplication routine defined on the semiring
(algorithm~\ref{algorithm:naiveMinMatrixMult}).

On a graph with nonnegative edge weights, it is trivial to create a
bijective problem on $(\times, \max)$, by letting $W'_{i,j} = e^{-
  W_{i,j} }$: thus $e^{-\left( W_{i,j} + W_{j,k} \right)} =
e^{-W_{i,j}} \times e^{-W_{j,k}}$, indicating that $+$ operations have
been converted to $\times$ operations. Likewise, since $e^{-x}$ is a
strictly decreasing function on $\mathbb{R}_+$, then $\min$ has become
$\max$ in the transformed space; an equivalent method $F'$ can be
made, which operates on the semiring $(\times, \max)$. $F'$ is then
used to process transformed inputs
(algorithm~\ref{algorithm:naiveMaxMatrixMult}). It is now trivial to
construct $G$, which performs the same operations as $F$, but where
$\max$ (equivalent to $\min$ in the original space) is replaced by $+$
operations (algorithm~\ref{algorithm:naiveMatrixMultPlus}). Lastly,
$H$ is constructed as an algorithm numerically equivalent to $G$ that
achieves greater speed by using any sub-cubic matrix multiplication
algorithm. Therefore, if $H'$ is called (\emph{i.e.}, $H$ is called on
scaled inputs) using a small collection of $p^*$ values, it is
possible to create a sequence of norms for each cell $i,j$ in the
result, and from that sequence it is possible to estimate the maximal
path lengths $E_{i,j}(I)$ on $(\times, \max)$. These estimates can
then be transformed back onto $(+, \min)$ by using the inverse
transformation $-\log(E_{i,j}(I)) \approx F(I)_{i,j}$. From there, it
is clear that $H$ can be chosen as any fast matrix multiplication
algorithm (to that end, this manuscript uses the $O(n^{2.807})$
Strassen matrix multiplication algorithm, as shown in
algorithm~\ref{algorithm:strassen}).

The following strategy for max-matrix multiplication is proposed:
First, compute the standard matrix multiplication
\begin{multline*}
C^{(p^*)} =\\
\left[
\begin{array}{cccc}
{\left(X_{0,0}\right)}^{p^*} & {\left(X_{0,1}\right)}^{p^*} & \cdots & {\left(X_{0,n-1}\right)}^{p^*}\\
{\left(X_{1,0}\right)}^{p^*} & {\left(X_{1,1}\right)}^{p^*} & & {\left(X_{1,n-1}\right)}^{p^*}\\
 \vdots & & \ddots & \\
{\left(X_{n-1,0}\right)}^{p^*} & {\left(X_{n-1,1}\right)}^{p^*} & & {\left(X_{n-1,n-1}\right)}^{p^*}\\
\end{array}
\right] \cdot\\
\left[
\begin{array}{cccc}
{\left(Y_{0,0}\right)}^{p^*} & {\left(Y_{0,1}\right)}^{p^*} & \cdots & {\left(Y_{0,n-1}\right)}^{p^*}\\
{\left(Y_{1,0}\right)}^{p^*} & {\left(Y_{1,1}\right)}^{p^*} & & {\left(Y_{1,n-1}\right)}^{p^*}\\
 \vdots & & \ddots & \\
{\left(Y_{n-1,0}\right)}^{p^*} & {\left(Y_{n-1,1}\right)}^{p^*} & & {\left(Y_{n-1,n-1}\right)}^{p^*}\\
\end{array}
\right]
\end{multline*}
for each $p^* \in P$ (via Strassen's algorithm or any other fast
matrix multiplication defined on the ring $(\times, +)$). Then the
estimate of the max-matrix multiplication at index $i,j$ is computed
via the sequence made from vector norms to the $p^*$: 
\[
E_{i,j}(I) = \mbox{\bf estimateMaxFromNormPowerSequence}\left( (~
C^{(p^*)}_{i,j} ~\vert~ \forall p^* \in P : \| u \|_{p^*}^{p^*} \geq
\tau ~) \right).
\]
The min-matrix multiplication results can be computed by transforming
back to the semiring $(+, \min)$, as described above. Thus, by
broadcasting into a small number of $L_{p^*}$ spaces, it is possible
to approximate a single matrix multiplication on the semiring $(+,
\min)$ (algorithm~\ref{algorithm:subcubicApproxMaxMatrixMult}), and
thereby approximate the APSP path lengths in $\log(n)$ fast matrix
multiplications. In this case (using Strassen multiplication), the
overall runtime achieved is in $O(n^{2.807} \log(n))$, and is faster
in practice than the $O(n^3)$ Floyd-Warshall algorithm. As mentioned
above, other fast matrix multiplication algorithms may be used in
place of Strassen's algorithm. For example, the Coppersmith-Winograd
algorithm would make the overall runtime $O(n^{2.36}
\log(n))$. Ignoring the runtime constants (which, it should be noted,
can significantly slow down the more advanced faster-than-naive matrix
multiplication algorithms in practice), when $n={10}^6$, $n^3 =
{10}^{18}$, whereas $n^{2.36} \log(n) = 2.88 \times {10}^{15}$.

\begin{algorithm}
  \caption{$F_{(APSP)}$: Naive matrix multiplication on the
    semiring $(+, \min)$. The inputs $X$ and $Y$ are both matrices
  in $\mathbb{R_+}^{n \times n}$. The $(+, \min)$ matrix
  multiplication between $X$ and $Y$ is returned. The algorithm runs
  in $O(n^3)$ time.}
  \label{algorithm:naiveMinMatrixMult}
  \begin{small}
    \begin{algorithmic}[1]
    \For{$i \in \{ 0, 1, \ldots n-1 \}$}
      \For{$j \in \{ 0, 1, \ldots n-1 \}$}
        \For{$k \in \{ 0, 1, \ldots n-1 \}$}
        \State $R_{i,j} \gets \min( R_{i,j}, X_{i,k} + Y_{k,j} )$
        \EndFor
      \EndFor
   \EndFor
   \Return $R$
   \end{algorithmic}
\end{small}

\end{algorithm}

\begin{algorithm}[t]
\caption{ $F'_{(APSP)}$: Naive matrix multiplication on the semiring
  $(\times, \max)$. The inputs $X'$ and $Y'$ are both matrices in
  ${[0,1]}^{n \times n}$. The $(\times, \max)$ matrix multiplication
  between $X'$ and $Y'$ is returned in $O(n^3)$. The algorithm runs in
  $O(n^3)$ time.}
\label{algorithm:naiveMaxMatrixMult}
\begin{small}
\begin{algorithmic}[1]
\For{$i \in \{ 0, 1, \ldots n-1 \}$}
  \For{$j \in \{ 0, 1, \ldots n-1 \}$}
    \For{$k \in \{ 0, 1, \ldots n-1 \}$}
      \State $R_{i,j} \gets \max( R_{i,j}, X'_{i,k} \times Y'_{k,j} )$
    \EndFor
    \EndFor
    \EndFor
\Return $R$
\end{algorithmic}
\end{small}
\end{algorithm}

\begin{algorithm}[t]
\caption{ { $G_{(APSP)}$: Naive matrix multiplication on the ring
    $(\times, +)$.} The inputs $X'$ and $Y'$ are both matrices in
  ${[0,1]}^{n \times n}$.  The $(\times, +)$ matrix multiplication
  between $X'$ and $Y'$ is returned. The algorithm, identical to
  $F'_{(APSP)}$ but with $\max$ operations replaced with $+$
  operations, runs in $O(n^3)$ time.}
\label{algorithm:naiveMatrixMultPlus}
\begin{algorithmic}[1]
\For{$i \in \{ 0, 1, \ldots n-1 \}$}
  \For{$j \in \{ 0, 1, \ldots n-1 \}$}
    \For{$k \in \{ 0, 1, \ldots n-1 \}$}
      \State $R_{i,j} \gets R_{i,j} ~+~ X'_{i,k} \times Y'_{k,j}$
    \EndFor
    \EndFor
    \EndFor
\Return $R$
\end{algorithmic}
\end{algorithm}

\begin{algorithm}[t]
\caption{ { $H_{(APSP)}$: Sub-Cubic Strassen matrix multiplication on
    the ring $(\times, +)$.} The inputs $X'$ and $Y'$ are both
  matrices in ${[0,1]}^{n \times n}$.  The $(\times, +)$ matrix
  multiplication between $X'$ and $Y'$ is returned. The algorithm
  produces the same result as $G_{(APSP)}$, but in $O(n^{2.807})$
  time.}
\label{algorithm:strassen}
\begin{small}
\begin{algorithmic}[1]
\State $\left[\begin{array}{cc}
    a & b\\
    c & d\\
  \end{array}
  \right] \gets X'$

\State $\left[\begin{array}{cc}
    e & f\\
    g & h\\
  \end{array}
  \right] \gets Y'$

\State $p1 \gets H_{(APSP)}(a,f-h)$\\
\State $p2 \gets H_{(APSP)}(a+b,h)$\\
\State $p3 \gets H_{(APSP)}(c+d,e)$\\
\State $p4 \gets H_{(APSP)}(d,g-e)$\\
\State $p5 \gets H_{(APSP)}(a+d,e+h)$\\
\State $p6 \gets H_{(APSP)}(b-d,g+h)$\\
\State $p7 \gets H_{(APSP)}(a-c,e+f)$\\
\State $R \gets \left[\begin{array}{cc}
    p5 + p4 - p2 + p6 & p1 + p2 \\
    p3 + p4 & p1 + p5 - p3 - p7 \\
  \end{array}
  \right]$\\
\Return $R$
\end{algorithmic}
\end{small}
\end{algorithm}

\begin{algorithm}[t]
\caption{ {$E_{(APSP)}$: Sub-Cubic approximate matrix multiplication
    on the semiring $(\times, \max)$.} Inputs $X'$ and $Y'$ are both
  matrices in ${[0,1]}^{n \times n}$. The $(\times, +)$ matrix
  multiplication between $X'$ and $Y'$ is returned. The algorithm
  approximates $F'_{(APSP)}$ in $O(n^{2.807})$ time.}
\label{algorithm:subcubicApproxMaxMatrixMult}
\begin{small}
\begin{algorithmic}[1]
\For{$p^* \in P$}
  \For{$i \in \{0, 1, \ldots n-1\}$}
    \For{$j \in \{0, 1, \ldots n-1\}$}
      \State $X''_{i,j} \gets {\left(X'_{i,j}\right)}^p{^*}$
      \State $Y''_{i,j} \gets {\left(Y'_{i,j}\right)}^p{^*}$ \Comment{(Create $I^{p^*}$; to prevent overflow, instead call $H_{(APSP)}'(X', Y', p^*)$ below)}
    \EndFor
  \EndFor

  \State $H^{(p^*)} \gets H_{(APSP})(X'', Y'')$ \Comment{(Perform fast operations in rings)}
\EndFor

\For{$i \in \{0, 1, \ldots n-1\}$}
  \For{$j \in \{0, 1, \ldots n-1\}$}
    \State $normPowerSeq \gets ~( H^{(p^*)}_{i,j} ~\vert ~ \forall p^* \in P : H^{(p^*)}_{i,j} \geq \tau )$
    \State $R_{i,j} \gets \mbox{\bf estimateMaxFromNormPowerSequence}\left( normPowerSeq \right)$ \Comment{(Aggregate results)}
  \EndFor
\EndFor
\Return $R$
\end{algorithmic}
\end{small}
\end{algorithm}

\subsection{Fast Sorting of $x_i + y_j$}
The fast rings approximation is also demonstrated on a second
well-known computer science problem: Sorting a list of all pairs
$x_i+y_j$ where $x$ and $y$ are two $n$-length lists, is a classic
problem in computer science for which no known algorithms achieve
runtime superior to the naive $O(n^2 \log(n))$ approach; furthermore,
this naive $O(n^2 \log(n))$ approach
(algorithm~\ref{algorithm:naiveXPlusY}) is the fastest known approach
that can also give the indices of $x$ and $y$, by generating all
tuples of the form $(x_i+y_j, i, j)$ and sorting them
lexicographically. Note that sorting all $n^2$ pairs $x_i + y_j$ would
require $O(n^2 \log(n))$ steps and retrieving the top $k$ values from
a max-heap would require inserting all $n^2$ values and then dequeuing
the top $k$ in $O(k \log(n))$ for an overall runtime in $O(n^2 + k
\log(n))$ where $k \leq n^2$.

An approximate solution can be achieved by discretizing to integer
values and then binning $x$ and $y$ (the binned counts of $x+y$ values
can be computed by convolving the binned $x$ and binned $y$ counts
with FFT convolution); however, this approximation is sensitive to the
discretization precision (in both accuracy and runtime) and yields
only the sorted values $x_i+y_j$ and not the corresponding indices
$i,j$~\cite{erickson:lower}. Here a novel numerical approximation to
sorting $x+y$ is outlined using the strategies above. The proposed
method can also be used to estimate the top $k$ values $x_i+y_j$ as
well as estimate the indices $i,j$ that produce them. As before with
the APSP problem, it is possible to draw an isomorphism between the
semirings $(+, \max)$ and $(\times, \max)$: $x'_i = e^{x_i}, y'_j =
e^{y_j}$ so that $e^{x_i + y_j} = e^{x_i} \times e^{y_j}$, indicating
the $+$ operation has been converted to a $\times$ operation.

Due to recent advances mentioned above regarding fast max-convolution,
it is tempting to find a similarity; however, max-convolution does not
directly solve the top $k$ values in $x'_i \times y'_j$: Given $z = x
~*_{\max}~ y$ (\emph{i.e.}, the max-convolution between $x$ and $y$)
the largest value in the max-convolution must be the largest value in
$x'_i \times y'_j$: $\max_m~ z_m = \max_i~ \max_j~ x'_i \times
y'_j$. But the second largest value in the max-convolution does not
necessarily belong to the top $k$ values of $x'_i \times y'_j$,
because the max-convolution at index $m$ gives the maximum value over
all positive diagonals for which $i+j=m$; if the second largest value
in $x'_i \times y'_j$ has the same $m=i+j$ value as the first largest
value chosen, then it will be obscured by first choice because $z_m =
\max_i~ u^{(m)}_i$, where $u^{(m)}$ is a vector that holds all
elements $x'_i \times y'_j$ along the positive diagonal $m=i+j$, and
$z$ only contains the maximum value of $u^{(m)}$, not the second
highest value, third highest value, \emph{etc.}.

It is trivial to see a naive sorting approach that simply performs
$\mbox{\bf sort}\left( ~( x'_0 \times y'_0, x'_0 \times y'_1, x'_0
\times y'_2, \ldots x'_1 \times y'_0, x'_1 \times y'_1, \ldots )~
\right)$; but this sorting method is already obfuscated by the clever
optimizations inherent to $O(n \log(n))$ sorting algorithms. For this
reason, a different algorithm $F$ is chosen; this $F$ algorithm is
equivalent to sorting all pairs $x'_i \times y'_j$, but the chosen
definition of $F$ eschews the complexity of sophisticated sorting
routines in favor of something simpler, although it is slower than the
naive approach of generating all $n^2$ $x'_i \times y'_j$ pairs and
sorting them with an arbitrary $O(n \log(n))$ algorithm. Essentially,
the algorithm is equivalent to finding the maximum value along each
positive diagonal, and then the maximum value over those maximum
values, which gives the next largest value in $x'_i \times y'_j$
(algorithm ~\ref{algorithm:naiveXTimesY}). The $u^{(m)}$ vectors
correspond to the positive-sloping diagonals in the following matrix:
\[
\left[
\begin{array}{ccccc}
x'_0 \times y'_0 & x'_0 \times y'_1 & x'_0 \times y'_2 & & x'_0 \times y'_{n-1} \\
x'_1 \times y'_0 & x'_1 \times y'_1 & x'_1 \times y'_2 & \cdots & x'_1 \times y'_{n-1} \\
x'_2 \times y'_0 & x'_2 \times y'_1 & x'_2 \times y'_2 & & x'_2 \times y'_{n-1} \\
 & \vdots & & \ddots \\
x'_{n-2} \times y'_0 & x'_{n-2} \times y'_1 & x'_{n-2} \times y'_2 & & x'_{n-2} \times y'_{n-1} \\
x'_{n-1} \times y'_0 & x'_{n-1} \times y'_1 & x'_{n-1} \times y'_2 & & x'_{n-1} \times y'_{n-1} \\
\end{array}
\right].
\]
And so $u^{(0)} = (x'_0 \times y'_0)$, $u^{(1)} = (x'_1 \times y'_0,
x'_0 \times y'_1)$, $u^{(2)} = (x'_2 \times y'_0, x'_1 \times y'_1,
x'_0 \times y'_2)$, $\ldots u^{(2 n - 2)} = (x'_{n-1} y'_{n-1})$. Once
the next highest remaining value is computed, that value is removed
from future consideration by executing the line
$u^{(m)}.remove(\psi_{m^*})$, which traverses through the list
$u^{(m)}$ and removes the first value matching $\psi_{m^*}$. Overall,
this algorithm runs in $O(n^3)$ steps.

\begin{algorithm}[t]
\caption{ { $F_{(X + Y)}$: Naive sorting of $x_i + y_j$.} Inputs $x$
  and $y$ are lists of length $n$.  The return value is a list of the
  top $k$ values of the form $x_i + y_j$. The algorithm runs in $O(n^2
  \log(n))$ time and $O(n^2)$ space, where $k \in \{1, 2, \ldots
  n^2\}$.}
  \label{algorithm:naiveXPlusY}
\begin{small}
\begin{algorithmic}[1]
\State $unsorted \gets (~)$
\For{$i \in \{0, 1, \ldots n-1\}$}
  \For{$j \in \{0, 1, \ldots n-1\}$}
    \State $unsorted.append(x_i + y_j)$
  \EndFor
\EndFor
\State $sorted \gets \mbox{\bf sort}(unsorted)$
\Return $(~ sorted[0], sorted[1], \ldots sorted[k-1] ~)$
\end{algorithmic}
\end{small}
\end{algorithm}

\begin{algorithm}[t]
\caption{ { $F'_{(X' \times Y')}$: Naive sorting of $x'_i \times
    y'_j$.} The inputs are two lists $x'$ and $y'$ of length $n$. The
  return value is a list of the top $k$ values of the form $x'_i
  \times y'_j$.  The algorithm runs in $O(n^2 + n k)$ time and
  $O(n^2)$ space, where $k \in \{1, 2, \ldots n^2\}$.}
\label{algorithm:naiveXTimesY}
\begin{small}
\begin{algorithmic}[1]
\For{$m \in \{0, 1, \ldots n-1\}$}
  \State $u^{(m)} \gets (x'_m \times y'_0, x'_{m-1} \times y'_1, x'_{m-2} \times y'_2, \ldots x'_0 \times y'_m)$
  \State $\psi_m \gets \max_i~ u^{(m)}$
\EndFor

\State $R \gets (~)$
\For{$\ell \in \{1, 2, \ldots k\}$}
  \State $m^* \gets \argmax_m \psi_m$
  \State $R.append(\psi_{m^*})$
  \State $u^{(m^*)}.remove(\psi_{m^*})$ \Comment{(Remove the value from $u^{(m)}$)}
  \State $\psi_{m^*} \gets \max_i~ u^{(m^*)}$
\EndFor
\Return $R$
\end{algorithmic}
\end{small}
\end{algorithm}

Next, the algorithm $G$ is created from $F$
(algorithm~\ref{algorithm:naiveXTimesYPlus}); note that when it is
desirable, only a subset of the $\max$ operations may be replaced by
$+$ operations. Finally, a novel data structure motivated by this fast
rings approximation is created for sequentially approximating and
removing a maximum from a collection of norms, the {\tt
  NormQueue}. This novel data structure is paired with fast
max-convolution to construct the faster-than-naive algorithm $H$. This
underscores that it is also possible to perform dynamic programming
while in these various $p^*$-norm spaces, and therefore use the result
of a computation on the semiring $(\times, \max)$ to subsequently
alter the program flow.

The {\tt NormQueue} works as follows: for some vector $u^{(m)}$,
assume a collection of norms to the $p^*$ is given
\begin{eqnarray*}
NQ & = & \left( NQ_1, NQ_2, NQ_3, \ldots \right)\\
& = & \left( \| u^{(m)} \|_{p^*_1}^{p^*_1}, \| u^{(m)} \|_{p^*_2}^{p^*_2}, \| u^{(m)} \|_{p^*_3}^{p^*_3}, \ldots \right)
\end{eqnarray*}
from which it is possible to estimate the maximum value in $\| u^{(m)}
\|_\infty \approx \alpha = \mbox{\bf
  estimateMaxFromNormPowerSequence}( \| u^{(m)} \|_{{p^*}_1}, \|
u^{(m)} \|_{{p^*}_2}, \| u^{(m)} \|_{{p^*}_3}, \ldots )$. If the
maximum element in $u^{(m)}$ occurred at index $i^* = \argmax_i
u^{(m)}_i$, then it is now possible to estimate the norms of the
vector ${u^{(m)}}' = \left( u^{(m)}_i | i \neq i^* \right)$ by
subtracting the estimated maximum $\alpha$ from each norm:
\begin{eqnarray*}
NQ' & = & \left( \| {u^{(m)}}' \|_{p^*_1}^{p^*_1}, \| {u^{(m)}}' \|_{p^*_2}^{p^*_2}, \| {u^{(m)}}' \|_{p^*_3}^{p^*_3}, \ldots \right)\\
& \approx & \left( \| u^{(m)} \|_{p^*_1}^{p^*_1} - \alpha^{p^*_1}, \| u^{(m)} \|_{p^*_2}^{p^*_2} - \alpha^{p^*_2}, \| u^{(m)} \|_{p^*_3}^{p^*_3} - \alpha^{p^*_3}, \ldots \right)\\
& = & \left( NQ_1 - \alpha^{p^*_1}, NQ_2 - \alpha^{p^*_2}, NQ_3 - \alpha^{p^*_3}, \ldots \right),
\end{eqnarray*}
proceeding inductively to iteratively estimate the maximum from the
collection of norms and then removing the estimated maximum from those
norms. Note that the {\tt NormQueue} can be used with only $O(1)$
norms, leading to approximate results, but an $O(1)$ runtime to pop
and reestimate the new max. While this data structure does accumulate
error (because small imperfections in early estimates of the maxima
may have ripple effects on later estimates), the larger values
experience less of this error because they are retrieved first
(limiting the ripple effect of errors accumulated to that point) and
because the norm sequence best summarizes larger values (they are not
the values that endure underflow).

\begin{algorithm}[t]
\caption{ { $G_{(X' \times Y')}$: Method on the ring $(\times, +)$,
    derived from $F'_{(X' \times Y')}$.} The inputs are two lists $x'$
  and $y'$ of length $n$.  The return value is analogous to the return
  value of $F'_{X \times Y}$, but using the ring $(\times, \max)$. The
  algorithm runs in $O(n^2 + n k)$ time and $O(n^2 + k)$ space.}
\label{algorithm:naiveXTimesYPlus}
\begin{small}
\begin{algorithmic}[1]
\For{$m \in \{0, 1, \ldots n-1\}$}
  \State $u^{(m)} \gets (x'_m + y'_0, x'_{m-1} + y'_1, x'_{m-2} + y'_2, \ldots x'_0 + y'_m)$
  \State $\sigma_m \gets \sum_i~ u^{(m)}$ \Comment{($\max$ operation replaced with $\sum$)}
\EndFor

\State $R \gets (~)$
\For{$j \in \{1, 2, \ldots k\}$}
  \State $m^* \gets \argmax_m \sigma_m$
  \State $R.append(\psi_{m^*})$
  \State $u^{(m^*)}.remove(\psi_{m^*})$ \Comment{(Remove the value from $u^{(m)}$)}
  \State $\sigma_{m^*} \gets \max_i~ u^{(m^*)}$
\EndFor
\Return $R$
\end{algorithmic}
\end{small}
\end{algorithm}

In order to derive some $H_{X' \times Y'}$ that is equivalent to but
more optimized than $G_{X' \times Y'}$, notice the fact that the
$\psi$ vector computed by $G_{X \times Y}$ is equivalent to a standard
convolution. One avenue of attack would be to compute $\psi$ for
different $p^* \in P$ (denoted $\sigma^{(p*)}$), and then aggregate
them to approximate the maximum (computed as the initial $\psi_m$ by
$F_{X' \times Y'}$). As noted above, the max-convolution will not
solve this problem alone, since the max-convolution discards the
second-highest value in each $u^{(m)}$ (once again, operations on the
semiring lose information in a manner that cannot be undone). However
in this case, it is desirable to keep the information in the
$p^*$-norm, so that the second-highest value in each $u^{(m)}$ leaves
some preserved signature. For this reason, the function $H$ is not
defined; instead, the $p^*$-norm aggregation is performed within the
same function $E$; instead of $E$ calling $H$ (as in the case of
max-convolution where $H$ is FFT and in max-matrix multiplication
where $H$ is Strassen-- or some other faster-than-naive-- matrix
multiplication), here $E_{X' \times Y'}$ is described \emph{ab initio}
so that $p^*$-norm information from the different calls to $H$ can be
shared (algorithm~\ref{algorithm:fastXTimesY}). By doing so, the
method can exploit the fact that if $\sigma^{(p^*)} = \sum_{i} {\left(
  u^{(m)} \right)}^{p^*}$, and so can recompute which value
$\sigma^{(p^*)}$ would take if its maximum ($\psi_{m^*}$) were removed
by simply subtracting out the term the maximum would contribute to the
norms: $\sum_{i} {\left( u^{(m)} \right)}^{p^*} -
{\left(\psi_{m^*}\right)}^{p^*}$. In practice, by using the {\tt
  NormQueue}. 

\begin{algorithm}[t]
\caption{ { $E_{(X' \times Y')}$: Fast approximate sorting of $x'_i
    \times y'_j$ operating on ring $(\times, \max)$.} The inputs are
  two nonnegative lists $x'$ and $y'$ of length $n$.  The return value
  is an list approximating the top $k$ values of the form $x'_i \times
  y'_j$. The algorithm runs in $O( (n+k) \log(n))$ time and $O(n+k)$
  space.}
\label{algorithm:fastXTimesY}

\begin{small}
\begin{algorithmic}[1]

\For{$p^* \in P$}
  \State $\sigma^{(p^*)} \gets (x')^{p^*} ~*~ {y'}^{p^*}$ \Comment{(Convolve with FFT)}
\EndFor

\State $q \gets maxHeap()$
\For{$m \in \{0, 1, \ldots n-1\}$}
  \State $\psi_m \gets \mbox{\bf estimateMaxFromNormPowerSequence}\left( (\sigma^{(p^*)}_m~ \vert ~ \forall p^* \in P : \sigma^{(p^*)}_m \geq \tau ) \right)$
  \State $q.insert( (\psi_m, m) )$
\EndFor
      
\State $R \gets (~)$
\For{$j \in \{1, 2, \ldots k\}$}
  \State $m^* \gets q.popMax()$
  \State $R.append(\psi_{m^*})$
  
  \For{$p^* \in P$}
    \State $\sigma^{(p^*)}_m \gets \sigma^{(p^*)}_m - {\left(\psi_{m^*}\right)}^{p^*}$ \Comment{(Remove the value from $u^{(m)}$)}
    \State $\psi_{m^*} \gets \mbox{\bf estimateMaxFromNormPowerSequence}\left( (\sigma^{(p^*)}_{m^*})~ \vert ~ \forall p^* \in P : \sigma^{(p^*)}_{m^*} \geq \tau ) \right)$
  \EndFor
\EndFor
\Return $R$
\end{algorithmic}
\end{small}
\end{algorithm}

Not only does this achieve a very fast approximation of the maximum
$k$ values in $x_i \times y_j$, $E_{X' \times Y'}$ can be called twice
and both calls can be used together to estimate the indices $i,j$ that
correspond to each value $x_i \times y_j$: this is achieved by first
noting that $E_{X' \times Y'}(X', Y')$ should give the same result
regardless of the order of the elements in $Y'$. Thus, if $Y'[::-1]$
denotes the reverse of $Y'$ (Python notation), then $E_{X' \times
  Y'}(X', Y') \approx E_{X' \times Y'}(X', Y'[::-1])$. Second, the
value $m^*$ (\emph{i.e.}, the positive diagonal from which each next
value is drawn) can be trivially added to the return value by simply
changing the line $result.append(\psi_{m^*})$ to $result.append(~ (
\psi_{m^*}, m^* ) ~)$. Therefore, by computing $E_{X' \times Y'}(X',
Y')$ and also computing $E_{X' \times Y'}(X', Y'[::-1])$, it is
possible to get estimates for the positive diagonal from which each
element was drawn. Let the positive diagonal for a given result value
(from calling $E_{X' \times Y'}(X', Y')$) be denoted $m_1^* = i+j$ and
let the negative diagonal from which that same result was drawn (from
calling $E_{X' \times Y'}(X', Y'[::-1])$) be denoted $m_2^* = i +
(n-1-j)$. It can therefore be seen that $m_1^* + m_2^* = 2 i + n-1$,
and so $i = \frac{m_1^* + m_2^* - n + 1}{2}$. Once $i$ is computed,
then $j = m_1^* - i$. These estimates can subsequently be verified by
comparing the approximation $\psi_{m^*}$ with the empirical value $x_i
\times y_j$ using the $i,j$ computed above; when both are close, then
the $i,j$ value is a reliable estimate. The ability to estimate the
indices in this manner is significant, because an existing
approximation that creates binned histograms for $X$ and $Y$ and then
uses those to compute the binned histogram of $X+Y$ (by convolving the
$X$ and $Y$ histograms with FFT)~\cite{erickson:lower} cannot be used
to estimate indices: doing so would require passing indices through
the FFT, which would require a set of integers to be stored for every
value in the FFT (rather than a complex floating point value) and
those sets will take on several values as the FFT recurses, including
the full set $\{0, 1, \ldots n\}$. As a result, passing indices
through the FFT cannot yet be accomplished in $o(n^2)$ time.

\section{Results} \label{sec:results}

\subsection{All-Pairs Shortest Paths in a Weighted Graph}
Here an elegant and standard $O(n^3)$ algorithm, the Floyd-Warshall
algorithm, for solving the APSP problem is compared against a simple,
novel method that was created using the fast rings approximation
strategy. This novel method (made quickly and without a great
expertise on the APSP problem) achieves a fairly good approximation
and outperforms the Floyd-Warshall algorithm, as shown in
table~\ref{table:apsp-runtimes}. Note that even with a simple and
fairly numerically unstable fast matrix multiplication algorithm (a
naive implementation of the Strassen algorithm), the approximation is
not only close to the exact value, it also often improves as the
problem size increases, because there is an increased chance of
finding an efficient path between two vertices. The index $(1,0)$
(arbitrarily chosen as the first non-trivial index-- index $(0,0)$
necessarily has a distance of $0$) of a single $n=512$ problem yields
an exact shortest path distance of $5.1955813$ and an approximate
shortest path distance $5.2130877$ (absolute error $<0.01751$).

\begin{table}
\footnotesize
\centering
\begin{tabular}{r|cccccc}
\hline
$n$ & 16 & 32 & 64 & 128 & 256 & 512 \\
\hline \\
Floyd-Warshall runtime & 0.01155 & 0.09297 & 0.7717 & 5.956 & 49.43 & 378.9 \\
Fast rings approximation runtime & 0.02527 & 0.09537 & 0.4572 & 2.594 & 15.31 & 103.5 \\
MSE & 0.04587 & 0.05395 & 0.03049 & 0.02767 & 0.02228 & 0.01207 \\
\hline
\end{tabular}
\caption{{\bf Runtime comparison between Floyd-Warshall and the fast
    rings approximation method.} The practical runtimes of the
  $O(n^3)$ Floyd-Warshall algorithm are compared to the proposed
  $O(M(n) \log(n))$ runtime $p^*$-norm rings approximation (in this
  case, using $p^*_{\max}=512$ (\emph{i.e.}, $|P|=18$) and using
  Strassen matrix multiplication, \emph{i.e.}, $M(n)=O(n^{2.807})$)
  for problems of various size. For each problem size $n$, $5$
  different problems were simulated (by randomly generating adjacency
  matrices where off-diagonal values were uniformly sampled floating
  point values $\in [1,100]$). The mean over these $5$ replicates for
  each $n$ is reported, as well as the mean squared error (MSE)
  between the exact shortest path weights and the approximate shortest
  path weights. Note that the speedup grows with $n$. The average
  square error is $< \pm 0.05$ for all problems considered.}
\label{table:apsp-runtimes}
\end{table}

\subsection{Sorting $x_i + y_j$}
\begin{figure}
\centering
\includegraphics[width=5in]{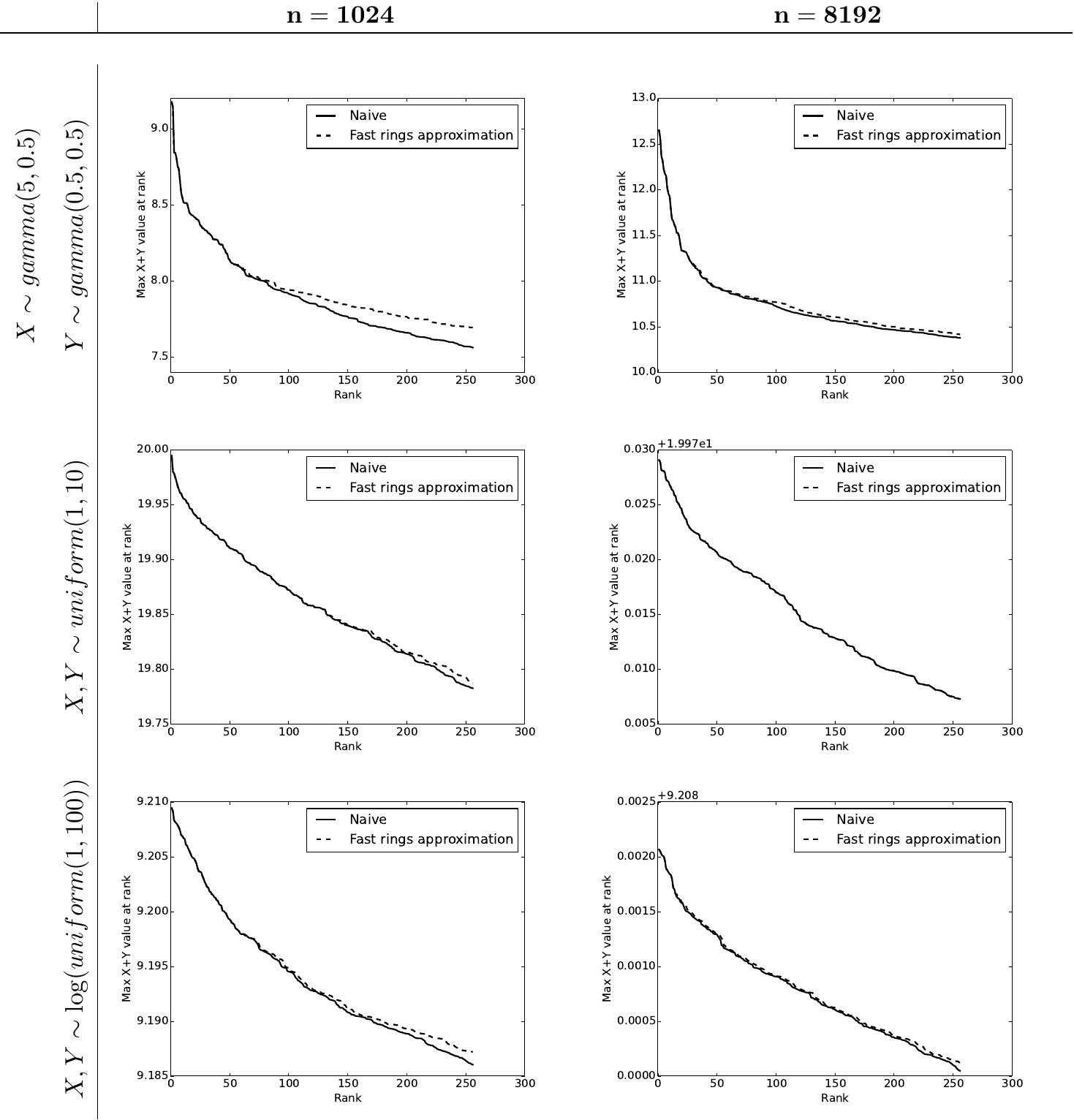}
\caption{{\bf Sorting $x_i + y_j$.} For problems of different size
  ($n=1024$ and $n=8192$) a single problem (\emph{i.e.} lists $x$ and
  $y$) is sampled from various distributions. On each problem, the
  naive $O(n^2 \log(n))$ approach is used to sort all possible $n^2$
  pairs and the top $k=256$ values are compared with the top values
  estimated with the fast rings approximation in $O(n \log(n))$ time
  (using $p^*_{\max}=4096$, \emph{i.e.}, $|P|=24$). Where the
  approximation is accurate, the indices $i,j$ that correspond to $x_i
  + y_j$ at every rank can be estimated with high fidelity
  (\emph{e.g.}, all $\frac{256}{256}$ top indices are correct for the
  uniform distributions with $n=8192$). When $n=8192$, the naive
  method required $873.2$ seconds and $\approx 15$GB of RAM, while
  fast approximation requires $0.7209$ seconds and memory usage was
  insignificant in comparison.}
\label{figure:xPlusY}
\end{figure}

Here the best known method (which is the naive $O(n^2 \log(n))$
algorithm) is demonstrated for finding the top $k$ values in $x_i +
y_j$, and compared to a novel $O(n \log(n) + k)$ algorithm based on the
fast rings approximation. Once again, the fast rings approximation
achieves a superior runtime (figure~\ref{figure:xPlusY}). Although the
error is significantly higher in this example (because errors are
accumulating in an iterative manner, which is not the case in
max-convolution and the APSP problem), not only is the runtime
superior, the space requirements are dramatically decreased (from
$>14$GB to memory usage in the low MBs due to a linear space
requirement), because the matrix of all $x'_i \times y'_j$ (or,
equivalently, $x_i + y_j$ in the original, un-transformed space) is
never actually computed.

\section{Discussion}\label{sec:discussion}
This manuscript has proposed novel methods for two open problems;
although the methods themselves are certainly of interest, most
exciting is that both were created without expertise even though both
problems (like max-convolution) have been subject to years of hard
work by the field. Furthermore, the methods themselves are fairly
simple (as was the case for fast max-convolution); other than code for
{\bf estimateMaxFromNormPowerSequence} described in
~\cite{pfeuffer:bounded} and implemented in an efficient, vectorized
manner with numpy, devising novel approximations is quite simple with
this general strategy: just as there are many problems to which this
strategy can be applied, there are many other variants of the
strategy. The general idea-- changing the problem into a spectrum of
rings, applying fast algorithms limited to rings, and then estimating
the true result from the aggregated spectrum-- could also be paired
with other soft-max functions that correspond well to the rings.

Even though the APSP approximation only estimates the path weights or
distances of the shortest path (as opposed to the path itself), there
may be situations where branch and bound can be used to compute the
paths in $o(n^3)$ time given knowledge about the optimal path lengths
(\emph{i.e.}, if the final paths are very efficient, then many edges
can be excluded wherever the cumulative distance sufficiently exceeds
the optimal path distance). Approximation error can even be worked
into this scheme by using a small buffer $\delta$ that prevents
bounding unless the weight is more than $\delta$ greater than the
approximation estimates. Likewise, an approximation may also be
suitable for cases from operations research where users are optimizing
over graphs, and thus may need to repeatedly estimate the APSP path
distances (in this case, it may even be possible to perform local
optimizations directly in the $p^*$-norm rings, which will be
continuous and differentiable). There may also be a strategy similar
to how indices can be estimated on the $x_i + y_j$ problem, which
could permit estimation of the destination of each edge taken by
matrix multiplication on $(\times, \max)$.

Furthermore, the largest errors appear to occur when the shortest path
between two vertices is long; this is because the $(\times, \max)$
variant of the problem operates in a transformed space, and a total
path weight of $20$ in the original space $(+, \min)$ corresponds to
$e^{-20}$ in the transformed space. For this reason, it is promising
to consider the application of scaling to keep the problem in a nicely
bounded range or even the possibility of using an alternative
transformation and then solving that problem on $(\times, \max)$. In
general, it may also be possible to solve $(+, \min)$ problems in a
similar manner without transforming to a $(\times, \max)$ formulation.

Considering the $x_i + y_j$ method, there may be other uses for the
{\tt NormQueue} data structure, the fast, approximate data structure
for keeping track of the largest remaining values in a collection
based on its norms (rather than storing the values themselves). There
may be other applications (\emph{e.g.}, in large-scale databases or
web search) where it is possible to cache a small collection of norms
(potentially with great efficiency via algorithms like FFT
convolution), but where caching the full list would be
intractable. Obvious applications would be similar to the $x_i + y_j$
problem, where a combinatorial effect makes caching results much more
difficult, and where algorithms such as FFT convolution can be used to
compute the norms on some combination of variables an order of
magnitude faster than if the norms were computed from scratch. This
sequence of norms can also be updated online in other ways (in
addition to the $popMax$ operation employed), such as adding values to
the queue (by adding in $x^{p^*}$ to the values stored at each $p^*$).

Regarding the $x_i + y_j$ problem itself, it would be very interesting
to see if the error of this preliminary algorithm could be improved by
performing $E_{X' \times Y'}(X', Y')$ and $E_{X' \times Y'}(X',
Y'[::-1])$ simultaneously (rather than serially, as was used for
estimating the indices $i,j$ in this manuscript). If both instances
proceed one index at a time, then an estimate of indices $i,j$ could
be computed before calling $popMax$; if the estimated indices are
accurate, then the exact value $x_i + y_j$ could replace the estimated
value when updating the {\tt NormQueue} (\emph{i.e.}, when subtracting
out the estimated max to the power $p^*$ from each $\| u
\|^{p^*}_{p^*}$), and such updates would introduce error much more
slowly (because, in an inductive manner, starting with high-quality
estimates of the initial maxima would yield to more accurate updating
of the {\tt NormQueue}, which would lead to higher quality estimates
of subsequent maxima).

This modularity of this fast rings approximation is a substantial
benefit: in the likely case that future research discovers alternative
methods for computing {\bf estimateMaxFromNormPowerSequence} (which in
this manuscript uses an $r=2$ projection), the accuracy or the
speed-accuracy tradeoff of this method would immediately
improve. Future developments in the conjectured error bound for the
$r=2$ projections will be of interest, as will error bounds for the
$r=3$ and $r=4$ projections, for which closed-form polynomial roots
can be computed. Just as the error dramatically improves when changing
from the $\| u \|_{p^*} \approx \| u \|_\infty$ approximation to the
$r=2$ projection (indeed, the error bound of the former depends on
$n$, the length of $u$, whereas the relative error bound conjectured
for the $r=2$ projection no longer depends on $n$), using $r=3$ or
even using different models of the norm may pose even greater
advantages. It is also important to note that although $r=3$ or $r=4$
models will almost certainly be slightly slower, they may also require
smaller $|P|$ sets to achieve the same error, thereby lowering the
runtime again.

This same modularity that lets new estimates of maxima be used easily
also allows faster algorithms $H$ on the corresponding ring space
(\emph{e.g.}, improved algorithms for matrix multiplication,
convolution, \emph{etc.}) to be used wherever such an algorithm can be
employed. This also means that sparse matrix multiplication could be
easily paired with the APSP approximation (although the advantages of
the fast rings approximation will almost certainly be mitigated for
such graphs). In cases where the dynamic range is large, the strategy
can also easily be used in the log-transform of the ring $(\times, +)$
(\emph{i.e.}, $(+, \log_+)$) to lower numerical error. High (variable)
precision numbers could be used with this strategy (introducing
computational complexity in each arithmetic operation, but allowing
for fewer $p^*$ to be used while attaining a high accuracy).

It would even be possible to compute a small number ($o(n)$ or even
$O(1)$) of exact solutions to sub-problems (\emph{e.g.}, for the APSP
problem, computing pairwise distances between a small number of vertex
pairs with Dijkstra's algorithm), and then use the relationship
between those exact values and the approximate values to build a
simple, affine model to correct numerical error (like the affine model
for correcting max-convolution results
from~\citet{pfeuffer:bounded}). This can be used to reduce bias and
substantially lower the MSE. Like using $p^*$-norm rings, that
strategy also generalizes to plenty of other problems. Also
reminiscent of the previous work on max-convolution is the ease of
parallelizing the approach (even coarse-grained parallelization),
because the $p^*$ rings can often be solved in parallel (\emph{e.g.},
the matrix multiplications for each $p^*$ in the APSP problem could be
trivially parallelized). The numerical error may also be reduced by
using more precise alternatives to Strassen's matrix multiplication
algorithm, because Strassen's algorithm is substantially less stable
than naive matrix multiplication. Fast but stable algorithms for
matrix multiplication will decrease $\tau$ for that problem, thereby
increasing the highest stable $p^*$ that can be used, and as a result
substantially lowering the numerical error.

It would also be interesting to investigate this approach on other
problems on semirings; the two problems discussed here (\emph{i.e.},
the APSP problem and finding the top $k$ values in $x_i + y_j$) were
chosen arbitrarily. This approximation method would likely be of
greatest utility on applications where little prior research exists
(in contrast with the APSP problem, for example).

\section{Availability}
Python code demonstrating these ideas is available at
\url{bitbucket.org/orserang/fast-semirings}.

\section{Acknowledgments}
I am grateful to Mattias Fr{\aa}nberg, Julianus Pfeuffer, Xiao Liang,
Marie Hoffmann, Knut Reinert, and Oliver Kohlbacher for their fast and
useful comments. O.S. acknowledges generous start-up funds from Freie
Universit\"{a}t Berlin and the Leibniz-Institute for Freshwater
Ecology and Inland Fisheries.


\end{document}